\documentclass[12pt]{iopart}

\usepackage{graphicx}
\usepackage{color}
\usepackage{url}

\newcommand{\ye}{$Y_{e}$~}
\newcommand{\alphan}{$(\alpha,n)$~}
\newcommand{\npa}{Nuclear Physics A}
\newcommand{\prc}{Physical Review C}

\newcommand{\prl}{Physical Review Letters}
\newcommand{\prep}{Physics Reports}
\newcommand{\apj}{ApJ}
\newcommand{\apjs}{ApJS}
\newcommand{\nat}{Nature}
\newcommand{\apjl}{ApJL}
\newcommand{\aap}{A\&A}
\newcommand{\araa}{Ann. Rev. Astron. \& Astrophy.}

\newcommand{\adndt}{At. Data Nucl. Data Tables}

\newcommand{\mnras}{MNRAS}
\newcommand{\physrep}{Physics Reports}
\usepackage{graphicx}
\usepackage{amssymb}

\begin{document}

\title{Impact of $(\alpha,n)$ reactions on weak r-process in neutrino-driven winds}

\author{J.~Bliss} 
\address{Institut f{\"u}r Kernphysik,
  Technische Universit{\"a}t Darmstadt, 64289 Darmstadt, Germany}
\author{A.~Arcones} 
\address{Institut f{\"u}r Kernphysik,
  Technische Universit{\"a}t Darmstadt, 64289 Darmstadt, Germany}
\address{GSI Helmholtzzentrum f\"ur Schwerionenforschung,
  Planckstr. 1, 64291 Darmstadt, Germany} 
\author{F. Montes} 
\address{National Superconducting Cyclotron Laboratory, Michigan
  State University, East Lansing, MI 48824, USA} 
\address{Joint
  Institute for Nuclear Astrophysics, http://www.jinaweb.org}
\author{J. Pereira} 
\address{National Superconducting Cyclotron
  Laboratory, Michigan State University, East Lansing, MI 48824, USA}
\address{Joint Institute for Nuclear Astrophysics,
  http://www.jinaweb.org}

\date{\today}

\begin{abstract}
After a successful core-collapse supernova, a neutrino-driven wind develops where it is possible to synthesize lighter heavy elements ($30<Z<45$). In the early galaxy, the origin of these elements is associated with the r-process and to an additional process. Here we assume the additional process corresponds to the weak r-process (sometimes referred to as) alpha-process taking place in neutrino-driven winds. Based on a trajectory obtained from hydrodynamical simulations we study the astrophysics and nuclear physics uncertainties of a weak r-process with our main focus on the \alphan reactions. These reactions are critical to redistribute the matter and allow it to move from light to heavy elements after nuclear statistical equilibrium freezes out. In this first sensitivity study, we vary all \alphan reactions by given constant factors which are justified based on the uncertainties of the statistical model and its nuclear physics input, mainly alpha optical potentials for weak r-process conditions. Our results show that \alphan rate uncertainties are indeed crucial to predict abundances. Therefore, further studies will follow to identify individual critical reactions. Since the nucleosynthesis path is close to stability, these reactions can be measured in the near future. Since much of the other nuclear data for the weak r-process are known, the reduction in nuclear physics uncertainties provided by these experiments will allow astronomical observations to directly constrain the astronomical conditions in the wind. 
\end{abstract}

%\pacs{26.30.-k, 26.30.Hj, 26.50.+x, 97.60.Bw}
\maketitle

% - - - - - - - - - - - - - - - - - - - - - - - - - - - - - - - - - - - - - - - -
\section{Introduction}
\label{sec:intro}
% - - - - - - - - - - - - - - - - - - - - - - - - - - - - - - - - - - - - - - - -

Heavy elements are produced by several nucleosynthesis processes and in various astrophysical sites. Elements heavier than iron are mainly formed by neutron-capture processes: slow neutron capture (s-process) and rapid neutron capture process (r-process). Other processes such as the alpha-process \cite{Meyer.etal:1992} or (also sometimes referred to as weak r-process \cite{Truran.Cowan:2000,Qian.Wasserburg:2007}), $\nu p$-process \cite{Pruet.etal:2006,Froehlich.etal:2006,Wanajo:2006}, p-process \cite{Arnould.Goriely:2003}, and i-process \cite{Cowan.Rose:1977,Hampel.etal:2016,Jones.etal:2016} may also contribute to the abundances of heavy elements but in less scale (at least for solar system abundances).

Understanding nucleosynthesis processes that operate under extreme conditions, such as the r-process, is a challenging problem because it involves scenarios with extreme and unknown astrophysical conditions and nuclear physics of the most neutron-rich nuclei. Even though the exact astrophysical environment for the r-process is not known, binary neutron star mergers (see e.g., \cite{Goriely.etal:2011, Korobkin.etal:2012, Wanajo.etal:2014}) and magnetorotationally driven supernovae (see e.g.,~\cite{Ji.etal:2016,Shibagaki.etal:2016,Nishimura.etal:2015}) appear to be possible scenarios to be considered. Despite the fast progress in the last years, the astrophysics and nuclear physics uncertainties are still relative large \cite{Arcones.etal:2016}. Recently, significant progress has been made on the uncertainty estimate of r-process abundance yields by nuclear physics properties such as masses, beta decay and neutron capture rates, and beta delayed neutron emission probabilities (see \cite{Mumpower.etal:2016,Martin.etal:2016} and references therein). Here we focus on the lighter heavy elements up to silver and their production in neutrino-driven winds after core-collapse supernovae. This is still challenging but also very exciting because most of the nuclei and nuclear reactions involved are relative close to stability. Therefore, it is possible that experimental work in the near future will significantly reduce the nuclear physics uncertainties of neutrino-driven winds. This will uniquely allow to use observation of the oldest stars to understand and constrain the extreme astrophysical conditions in the wind where the lighter heavy elements are synthesized \cite{Hansen.etal:2014}. 

There is evidence that at least at early times in the history of the Galaxy most of the abundances of heavy elements were created by the r-process and a still-to-be-decided nucleosynthesis process (or processes) \cite{Qian.Wasserburg:2007, Qian.Wasserburg:2008, Hansen.etal:2014} which at times has been referred to as LEPP (lighter element primary process) \cite{Travaglio.etal:2004,Montes.etal:2007,Arcones.Montes:2011}.  It is possible that this process is one of the listed above. In this paper we focus on the weak r-process that occurs in slightly neutron-rich neutrino-driven winds after successful core-collapse supernova explosions as a candidate for the creation of the lighter heavy elements. This process is also known as alpha-process because alpha-capture reactions are the underlying nucleosynthesis driver \cite{Hoffman.etal:1997}.

Recent core-collapse supernova simulations \cite{Huedepohl.etal:2010, Arcones.Thielemann:2013, Roberts.etal:2012,MartinezPinedo.etal:2012} indicate that neutrino-driven winds are proton rich or only slightly neutron rich. The exact neutron-richness and conditions of the neutrino-driven wind are still uncertain, but they are probably enough to synthesize lighter heavy elements (Sr to Ag) \cite{Arcones.Montes:2011, Arcones.Bliss:2014}. Here we will shortly discuss how the uncertainty on the neutron-richness impacts the abundances (for more details see \cite{Arcones.Thielemann:2013, Arcones.Bliss:2014}) and focus further on identifying key reactions that need to be measured. Slightly neutron-rich neutrino-driven winds are characterized by a nucleosynthesis path not far from stability where beta decays are much slower than the fast wind expansion. When matter starts expanding from the proto-neutron star, the nuclear statistical equilibrium (NSE) can produce nuclei up to $Z\sim40$. After the initial NSE phase, several reactions, that are faster than beta decays, keep moving matter from light to heavy nuclei and thus have an impact on the abundances. These reactions include $(\alpha,n)$, $(p,n)$, $(\alpha,\gamma)$, $(p,\gamma)$, as suggested in Ref.~\cite{Woosley.Hoffman:1992}. Ref.~\cite{Sasaqui.etal:2005} also investigated $(\alpha,n)$ reactions in context of the r-process. How uncertain are these reactions? What is the impact of such uncertainty on the abundances? Refs.~\cite{Pereira.Montes:2016,Mohr:2016} made first efforts to understand theoretical $(\alpha,n)$ reaction rate uncertainties. In this paper, we further study  $(\alpha,n)$ reactions: their uncertainties and impact on the nucleosynthesis.

In absence of experimental information, we use the statistical Hauser-Feshbach \cite{Hauser.Feshbach:1952} model (TALYS version 1.6 \cite{Talys1.6}) to explore the theoretical uncertainty of the $(\alpha,n)$  rates. The theoretical uncertainty is mainly due to alpha optical potentials for the temperatures relevant in the weak r-process \cite{Pereira.Montes:2016, Mohr:2016}. This investigation is used to estimate uncertainty factors that we use for a sensitivity study of the impact on abundances. Our results clearly show that  $(\alpha,n)$ reactions are critical to redistribute matter among different elements during the nucleosynthesis in neutron-rich neutrino-driven winds. Therefore, more detailed studies are required to identify individual critical $(\alpha,n)$  reactions.

The paper is organized as follow. The nucleosynthesis network is introduced in Sect.~\ref{sec:netw}. A general overview about the nucleosynthesis in neutron-rich winds is presented in Sect.~\ref{sec:nuc}. Astrophysical uncertainties and their impact on the nucleosynthesis are shortly included in Sect.~\ref{sec:astro_uncer}. In Sect.~\ref{sec:an_uncer}, we discuss and estimate the nuclear physics uncertainties of the $(\alpha,n)$  reactions. Their impact on the abundances is shown in Sect.~\ref{sec:results}, where we also compare to the impact of astrophysical uncertainties and to observations. We summarize and conclude in
Sect.~\ref{sec:summary}. 

% - - - - - - - - - - - - - - - - - - - - - - - - - - - - - - - - - - - - - - - -
\section{Nucleosynthesis network and reaction rates}
\label{sec:netw}
% - - - - - - - - - - - - - - - - - - - - - - - - - - - - - - - - - - - - - - - -

Our nucleosynthesis calculations are performed with the WINNET reaction network \cite{Winteler:2012, Winteler.etal:2012}. We consider 4412 neutron- and proton-rich nuclei as well as stable ones from H to Ir. The nuclear reaction rates are taken from JINA ReaclibV2.0 \cite{Cyburt.etal:2010} with the exception of the $(\alpha,n)$ rates and their inverse for all isotopes with $26<Z<45$, which are calculated with the reaction code TALYS~1.6 \cite{Talys1.6} (see Sect.~\ref{sec:an_uncer} for more details). The $(\alpha,n)$ rates are not fitted following the Reaclib prescription but interpolated from the values obtained directly from TALYS. This prevents having artificial divergences from the calculated rates which can influence our sensitivity study. The theoretical weak interaction rates are the same as in Ref.~\cite{Froehlich.etal:2006}.

The nucleosynthesis calculations start at a temperature around $T \approx 10~\mathrm{GK}$, thus assuming nuclear statistical equilibrium (NSE) until the temperature drops below $T=8~\mathrm{GK}$. The weak rates are considered from 10~GK, thus allowing $Y_e$ to evolve during the NSE phase. We include neutrino reactions on nucleons \cite{Froehlich.etal:2006}. Neutrino energies and luminosities, which are also used as network input parameters, are consistent with the initial \ye selected in our calculations (see \cite{Qian.Woosley:1996} for their relationship). Neutrons and protons dominate the initial abundances and its ratio is given by the initial $Y_{e}$. Heavy nuclei form as temperature and density decrease. 

% - - - - - - - - - - - - - - - - - - - - - - - - - - - - - - - - - - - - - - - -
\section{Nucleosynthesis in neutrino-driven winds}
\label{sec:nuc}
% - - - - - - - - - - - - - - - - - - - - - - - - - - - - - - - - - - - - - - - -

The nucleosynthesis evolution in neutron-rich winds has been extensively discussed especially for r-process \cite{Hoffman.etal:1997, Takahashi.etal:1994, Meyer.etal:1992}, but also for the weak r-process \cite{Arcones.Montes:2011, Arcones.Thielemann:2013, Arcones.Bliss:2014}. Even if both processes consist of neutron captures away from stability, the conditions are significantly different: in the weak r-process the neutron-to-seed ratio is very small, $Y_n/Y_\mathrm{seed} \lesssim 10^{-2}$, compared to the r-process $Y_n/Y_\mathrm{seed} > 100$. This results in an evolution close to or even along stability for the weak r-process and thus much longer beta decays of the nuclei involved.  This is a critical point since beta decays are much slower than the expansion time scale and therefore cannot be the main mechanism to move matter towards higher $Z$, as in the r-process.

For completeness, we summarize the nucleosynthesis evolution of the weak r-process using a trajectory obtained from spherically symmetric hydrodynamic simulations of neutrino-driven winds~\cite{Arcones.etal:2007}. We choose the trajectory ejected 9~s after bounce and an electron fraction $Y_e =0.47$, keeping the entropy ($S\approx 86$~k$_B$/nuc) and expansion time scale ($\tau =11$~ms, see \cite{Qian.Woosley:1996} for definition) as given by the simulation. After the initial NSE phase various reactions fall out of equilibrium and become important for the final redistribution of matter. Since the wind expansion is faster than beta decays, charged-particle reactions are the ones moving matter towards higher $Z$; these reactions include: $(\alpha,\gamma)$, $(\alpha,n)$ \footnote{In this paper  $(\alpha,n)$ refers also to alpha captures with the emission of one or more neutrons:  $(\alpha,\times n)$, with $\times =$~1,~2,~3.}, $(p,\gamma)$, and $(p,n)$.

\begin{figure}
 \centering 
 \includegraphics[width=0.75\linewidth,angle=0]{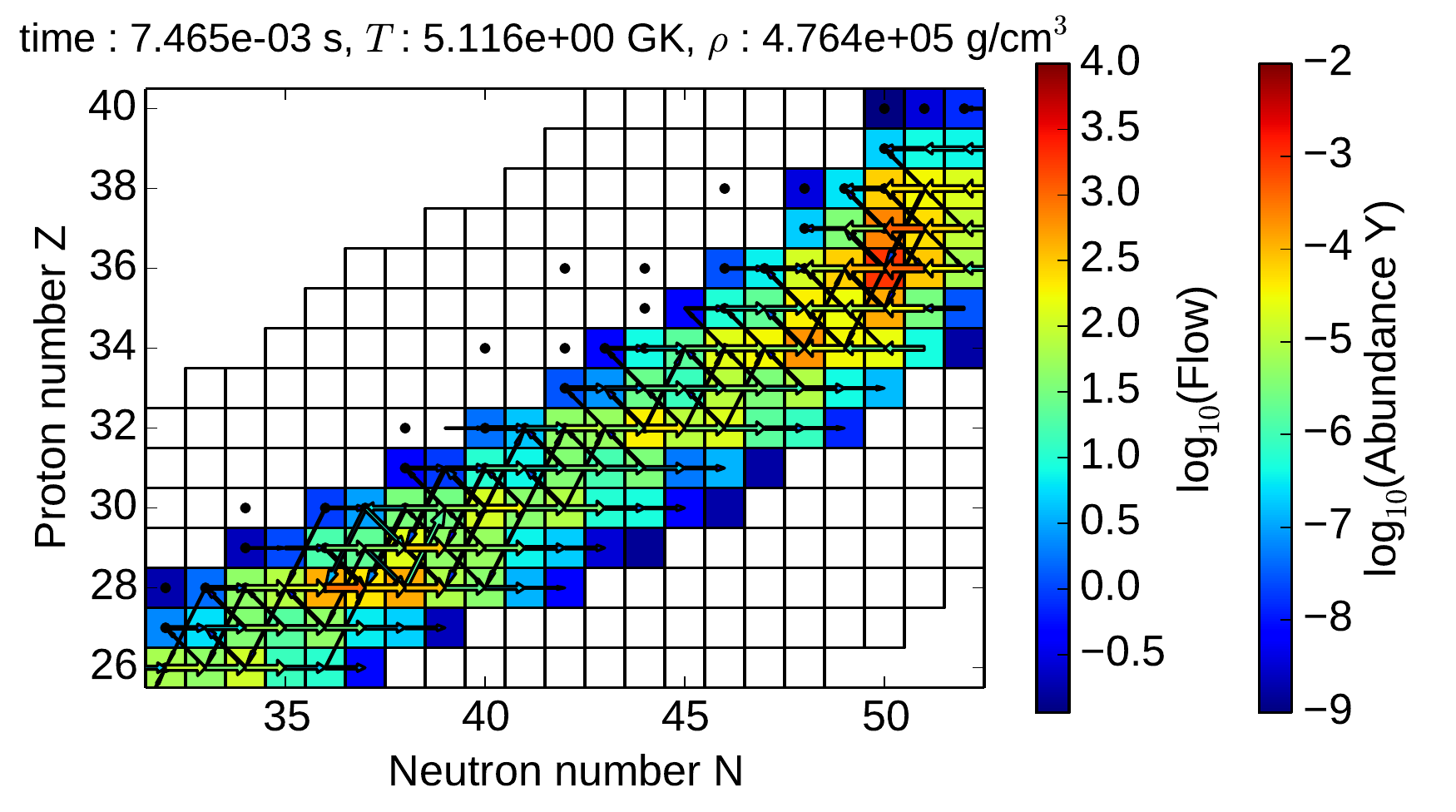}\\%{./plots/FlowYe47T5GK.eps}
 \includegraphics[width=0.75\linewidth,angle=0]{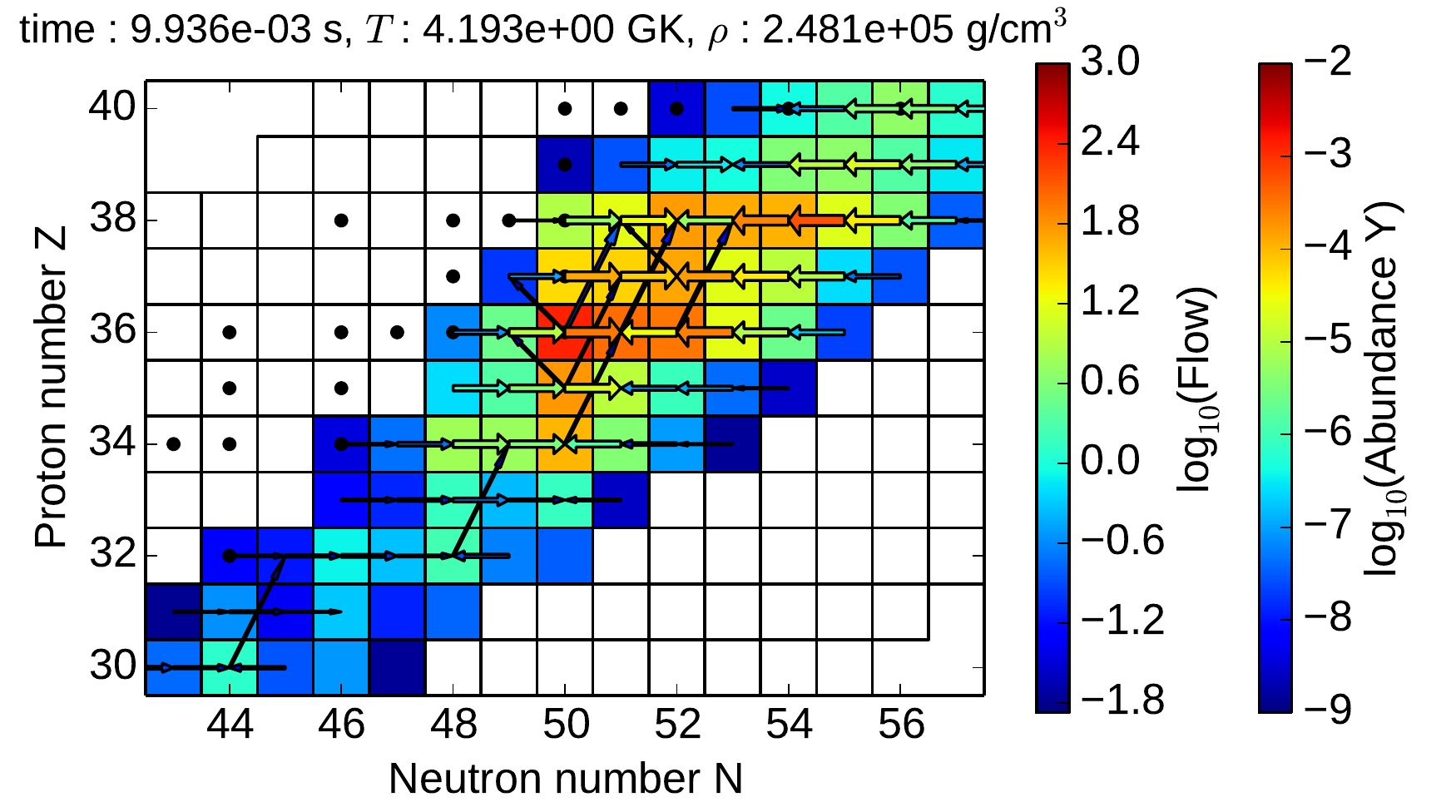}\\%{./plots/FlowYe47T4GK.eps}
 \includegraphics[width=0.75\linewidth,angle=0]{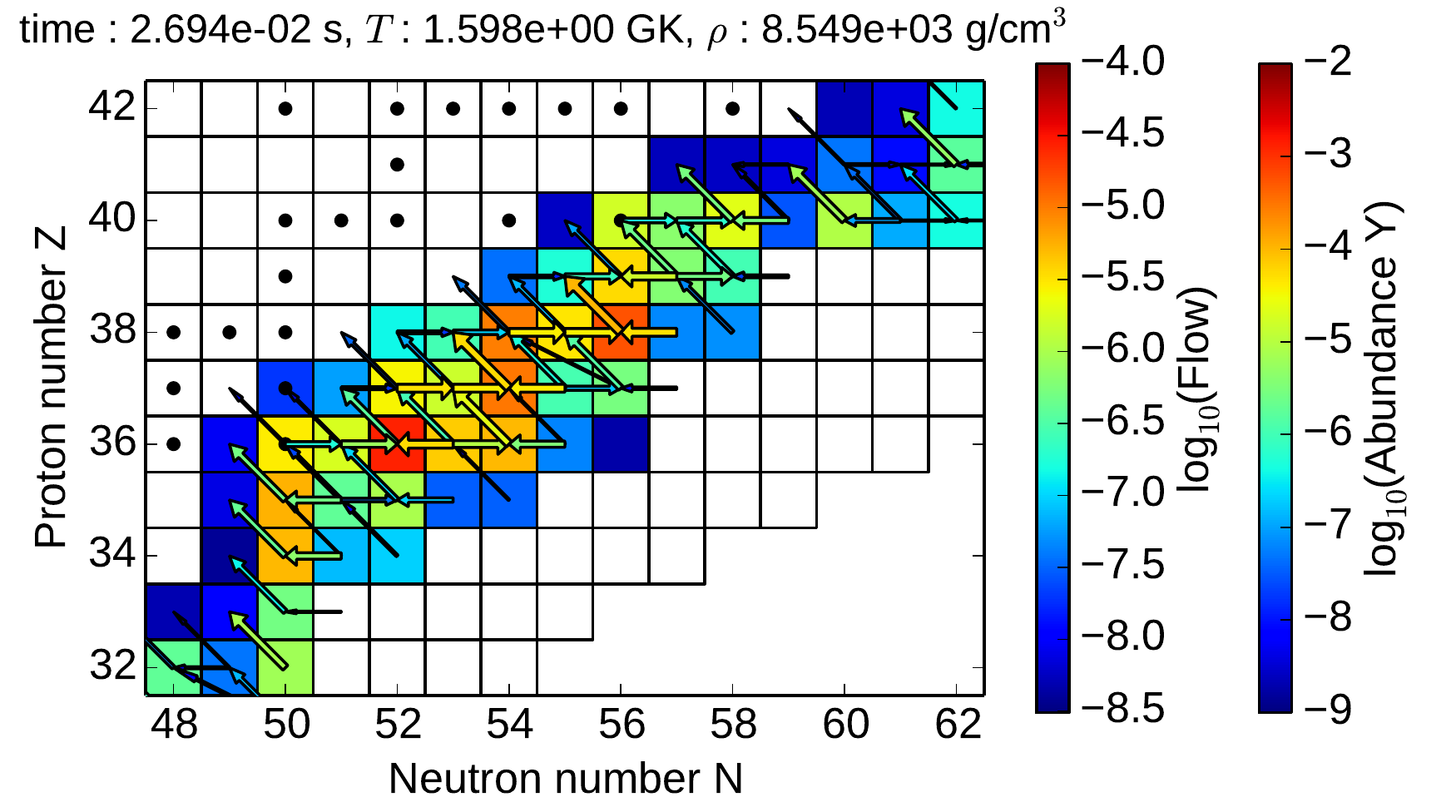}%
 \caption{The arrows show the flow of the different reactions. The colors and sizes of the arrows are proportional to the flows. The abundances are shown by different colors and stable nuclei are indicated by black dots.}
 \label{fig:Flow}
 \end{figure}

In order to analyze this important phase of the weak r-process, one can look at the reaction flows. The flow between two nuclei $i$ and $j$ is defined as
\begin{equation}
F_{ij} \equiv \dot{Y}(i \rightarrow j) - \dot{Y}(j \rightarrow i),
\end{equation}
where $\dot{Y}(i \rightarrow j)$ describes the change in abundance of nucleus $i$ due to all reactions connecting nucleus $i$ with nucleus $j$. Figure~\ref{fig:Flow} illustrates the nucleosynthesis evolution at different temperatures. At $T \approx 5.1$~GK (top panel in Fig.~\ref{fig:Flow}) the nucleosynthesis path has already reached the Sr, Y, Zr region. In every isotopic chain the nucleosynthesis path is given by $(n,\gamma)-(\gamma,n)$ equilibrium. Matter reaches higher $Z$ by $(\alpha,n)$ and $(p,n)$ reactions but $(n,\alpha)$ reactions still occur and carry some matter back to lighter nuclei. Around $T \approx 4.2$~GK (middle panel in Fig.~\ref{fig:Flow}) $(p,n)$ reactions are less important, while $(\alpha,n)$ reactions keep moving mater towards heavier nuclei. Note that besides increasing the proton number $Z$ by two units, the specific isotope resulting from a given ($\alpha$,$n$) channel depends on the number of neutrons emitted in that channel. 
However, as long as $(n,\gamma)-(\gamma,n)$ reactions are in equilibrium and much faster than ($\alpha$,$n$), the isotopic distribution within a given element is ultimately defined by the temperature, neutron density, and neutron separation energy~\cite{Fowler.etal:1967}. 
For typical weak r-process temperatures, the $(\alpha,1n)$ channel dominates the $(\alpha,n)$ reaction flux \cite{Pereira.Montes:2016}.
In agreement with \cite{Woosley.Hoffman:1992}, we find that in neutron-rich winds $(\alpha,n)$ reactions on Zn, Ge, Se, and Kr are keys to shift matter to heavier isotopes. 

\begin{figure}
\centering 
\includegraphics[width=0.7\linewidth,angle=0]{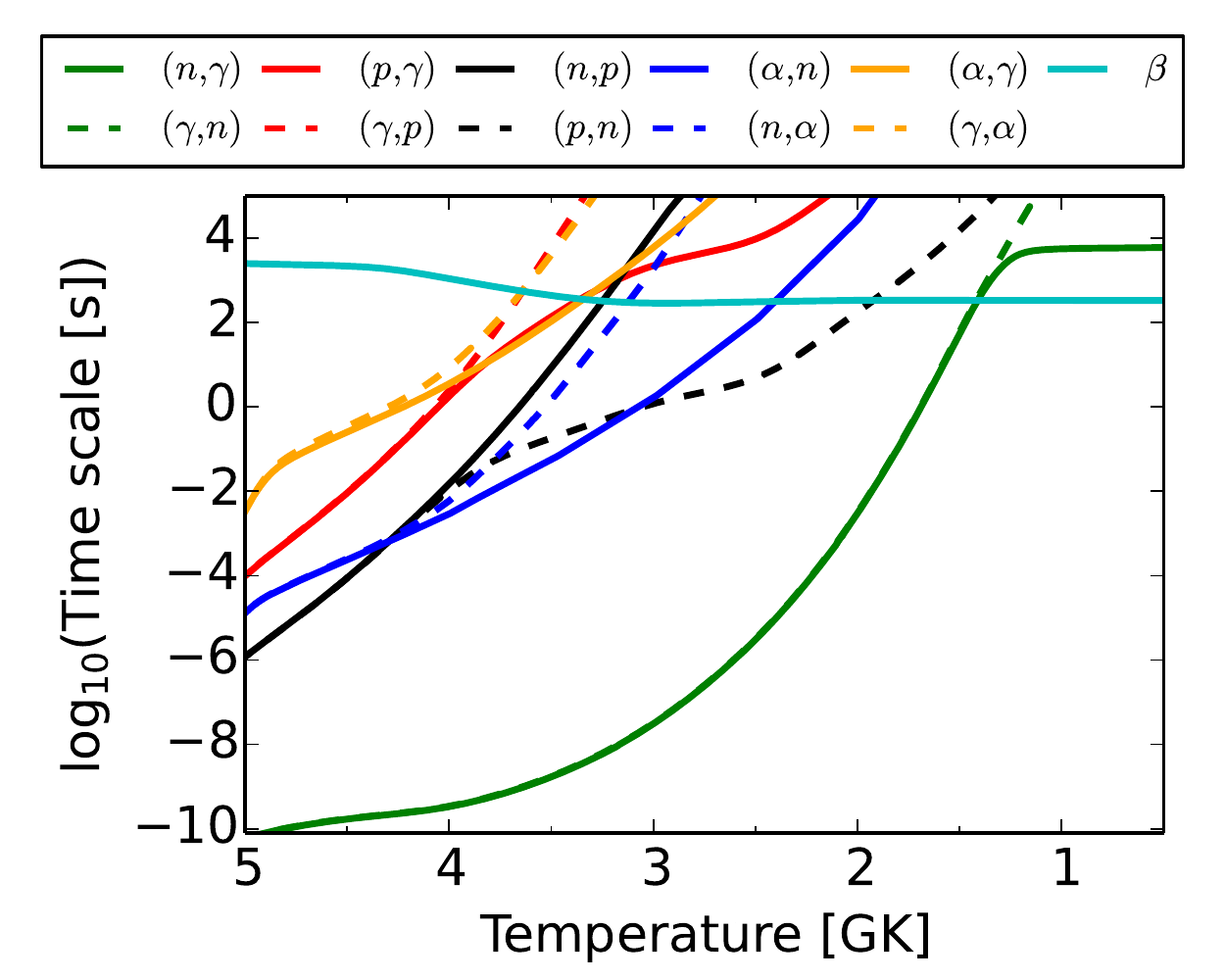}\\
\caption{Averaged time scales of important reactions as a
  function of temperature. This correspond to the trajectory ejected
  9~s after bounce with an initial electron fraction of \ye =0.47.}
\label{fig:Timescale}
\end{figure}

The description of the nucleosynthesis can be completed by considering the averaged time scales,  $\langle \tau_x \rangle$, of the most relevant reactions calculated as follows:
\begin{equation}
\frac{1}{\langle \tau_x \rangle} = \frac{\sum_{Z,A} \lambda_x(Z,A) Y(Z,A)}{\sum_{Z,A} Y(Z,A)},
\label{eq:tau}
\end{equation}
where $\lambda_{x} (Z,A)$ describes the reaction rate of process $x$ on nucleus $(Z,A)$ and $Y(Z,A)$ is the abundance of the nucleus.        
The averaged time scales for $Z=26-45$ are presented in Fig.~\ref{fig:Timescale} versus decreasing temperature. The fastest reactions are $(n,\gamma)$ and $(\gamma,n)$ which stay in equilibrium until the temperature drops below $T \approx 1.5~\mathrm{GK}$. For temperatures above $T \approx 4.2$~GK, $(p,n)$ reactions are the fastest charged-particle reactions. Note that in the fluxes (Fig.~\ref{fig:Flow}), $(p,n)$ reactions and beta decays are represented in the same way, however in the time scale figure it is clear that the driving reactions are $(p,n)$ and not beta decays for $T >1.9$~GK. Below $T \approx 4.2$~GK, $(\alpha,n)$ reactions become faster than $(p,n)$ reactions and they determine the nucleosynthesis evolution until the temperature drops down to $T \approx 3.3$~GK. At low temperatures (bottom panel in Fig.~\ref{fig:Flow}), the evolution is driven by beta decays and neutron captures instead of charged-particle reactions. There is a temperature range for which a type of reaction is important and this  depends on the astrophysical conditions. For example, for $Y_e=0.45$ the $(\alpha,n)$ reactions are already very important around $T \sim 4.5$~GK. For only slightly neutron-rich conditions ($Y_e \sim 0.49$), the nucleosynthesis path stays close to stability and $(p,\gamma)$, $(\alpha,\gamma)$ become more important to reach heavier nuclei, while \alphan reactions play a minor role.

% - - - - - - - - - - - - - - - - - - - - - - - - - - - - - - - - - - - - - - - -
\section{Astrophysics uncertainties}
\label{sec:astro_uncer}
% - - - - - - - - - - - - - - - - - - - - - - - - - - - - - - - - - - - - - - - -

The nucleosynthesis evolution described before is based on given astrophysical conditions that change during the evolution after the explosion and also depend on the supernova progenitor~\cite{Arcones.etal:2007, Arcones.Janka:2011}. The neutrino cooling of the neutron star leads to the evolution of the wind and the consequent variations of wind parameters: expansion time scale, entropy, and electron fraction \cite{Qian.Woosley:1996, Hoffman.etal:1997}. A fully self-consistent study of the importance of specific reactions on the abundances would require to analyze all possible astrophysical conditions, which is beyond the scope of the present work. Here we propose a first approach in which we explore different nucleosynthesis evolutions (i.e., paths along the nuclear chart) by varying the electron fraction. In this way, one can determine under which evolution a reaction is important. We have decided to vary $Y_e$ because this is the quantity that is more uncertain from the hydrodynamical simulations. Moreover, a variation of entropy and expansion time scale is not full consistent with the simulations (for a sensitivity study to entropy and expansion time scale see e.g.,~\cite{Hoffman.etal:1997,Otsuki.etal:2000,Arcones.Bliss:2014}).

% These parameters determine the key nucleosynthesis quantity: the neutron-to-seed ratio. In order to investigate different nucleosynthesis evolutions and thus estimate the astrophysical uncertainty, one can try all possible variations of the wind parameters and group together the ones resulting in the same neutron-to-seed ratio. Here we reduce the number of calculations by fixing two of the wind parameters, namely the entropy and the expansion time scale, and varying the third one ($Y_{e}$) \cite{Hoffman.etal:1997}. This still allows to investigate the influence of different neutron-to-seed ratios on the wind nucleosynthesis and relevant reactions. Similar results are obtained when varying the other wind parameters \cite{Arcones.Bliss:2014}. 

\begin{figure}
\centering 
\includegraphics[width=0.75\linewidth,angle=0]{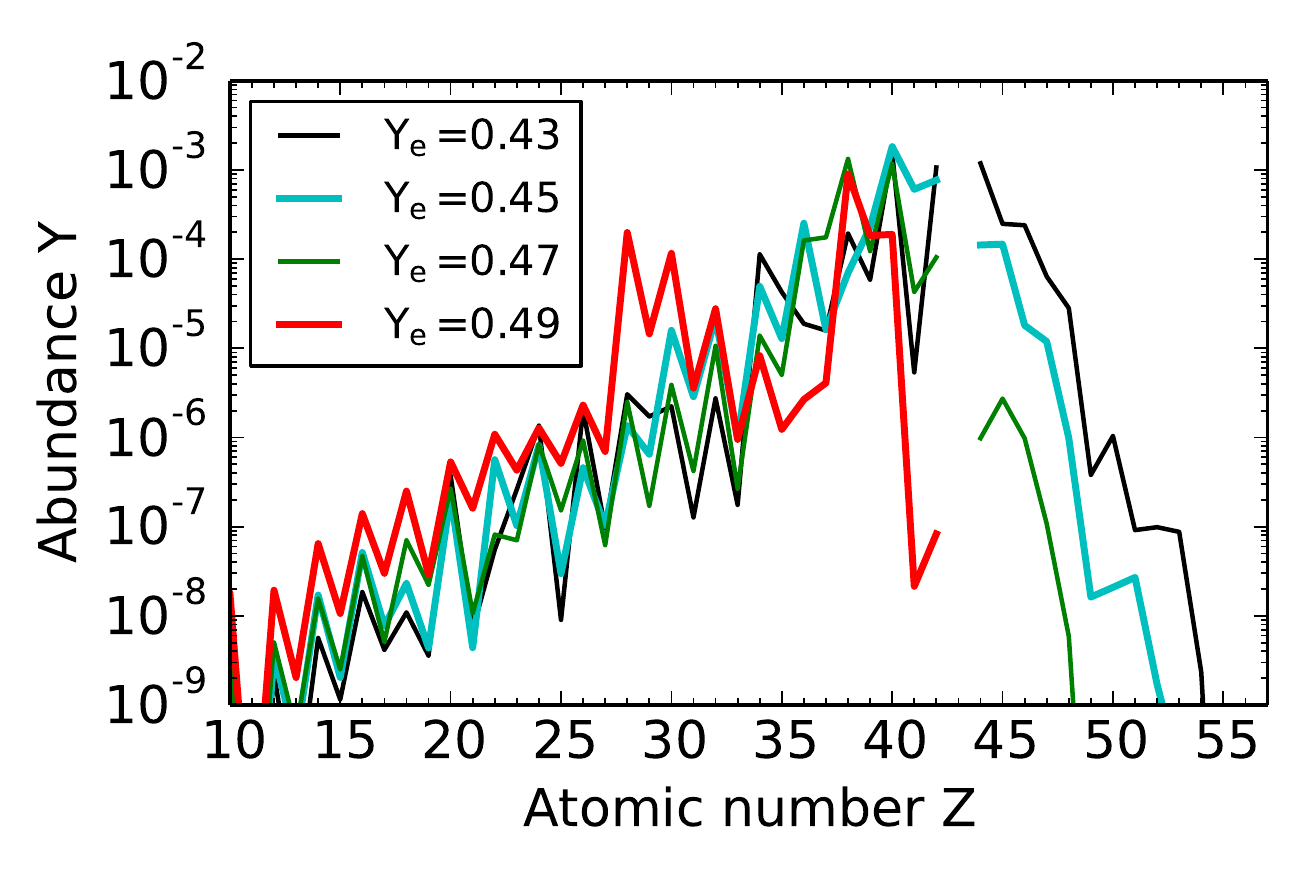}\\
\caption{Elemental abundances, i.e., $Y(Z)= \sum_{A} Y(Z,A)$, for the 9~s trajectory and different $Y_e
  =$~0.43, 0.45, 0.47, and 0.49.}
\label{fig:diffYe}
\end{figure}

The elemental abundances resulting from variations of the initial $Y_e$ are shown in Fig.~\ref{fig:diffYe}. In neutron-rich winds $(0.4 \lesssim Y_{e} \lesssim 0.5)$, a weak r-process \cite{Truran.Cowan:2000} can produce the lighter heavy elements, including Sr, Y, and Zr up to (possibly) Ag \cite{Qian.Wasserburg:2007,Wanajo:2013, Arcones.Bliss:2014,Wanajo.etal:2011}. When the electron fraction decreases and the path moves farther away from stability, the general trend is that heavier nuclei are reached. Moreover, variations of the nucleosynthesis evolution (i.e., different $Y_e$) lead to different patterns \cite{Arcones.Bliss:2014}. 
It is interesting to understand the different patterns for Sr, Y, and Zr since these variations are also observed in the elemental abundance of old stars \cite{Sneden.etal:2008,Hansen.etal:2013,Hansen.etal:2014,Roederer.etal:2014b}. In our calculations Sr, Y, and Zr show three typical patterns (see Fig.~\ref{fig:diffYe}):
\begin{enumerate}
  \item $Y($Sr$)< Y($Y$)<Y($Zr$)$,
  \item $Y($Sr$) \lesssim Y($Zr$)$ and $Y($Sr$)>Y($Y$)$,
  \item $Y($Sr$)> Y($Y$)>Y($Zr$)$.
\end{enumerate}
The third case occurs when the electron fraction is close to 0.5, then the path moves along stability and  Sr, Y, and Zr are the tail of the abundances. For lower $Y_{e}$, the abundances reach heavier elements and this explains the trend of $Y($Sr$)\lesssim Y($Zr$)$.

% - - - - - - - - - - - - - - - - - - - - - - - - - - - - - - - - - - - - - - - -
\section{\alphan reaction rate uncertainties}
\label{sec:an_uncer}
% - - - - - - - - - - - - - - - - - - - - - - - - - - - - - - - - - - - - - - - -

\begin{figure}
\centering
\includegraphics[width=0.8\linewidth,angle=0]{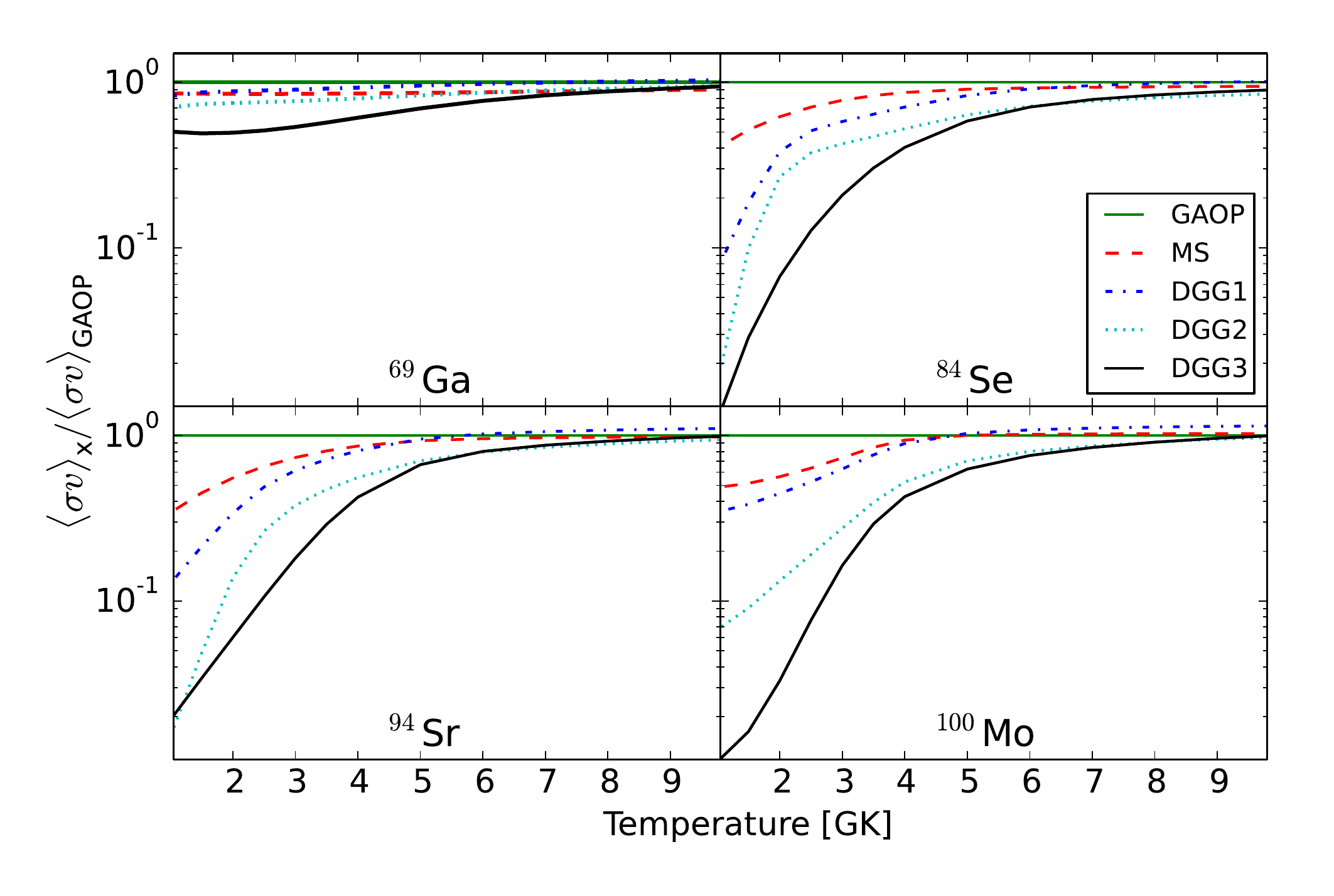}
\caption{Theoretical $^{69}\mathrm{Ga}(\alpha,n)$$^{72}$As, $^{84}\mathrm{Se}(\alpha,n)$$^{87}$Kr, $^{94}\mathrm{Sr}(\alpha,n)$$^{97}$Zr, and $^{100}\mathrm{Mo}(\alpha,n)$$^{103}$Ru reaction rates using the alpha optical potentials: global alpha optical potential (GAOP) \cite{Talys1.6,Watanabe:1958}, phenomenological fit of McFadden and Sachtler (MS) \cite{McFadden.Sachtler:1966}, three different versions of the model of Demetriou-Grama-Goriely (DGG1-3) \cite{Talys1.6,Demetriou.etal:2002} (the other nuclear inputs are determined from the default set of sources given in ~\cite{Pereira.Montes:2016}, with the exception of masses, which were taken from Ref.~\cite{Audi.etal:2003} if available, or from the FRDM mass model~\cite{Moeller.etal:1995} otherwise). The reaction rates are normalized to the ones calculated with the GAOP model.}
\label{fig:uncAOP}
\end{figure}

\begin{figure}
\centering
\includegraphics[width=0.49\linewidth,angle=0]{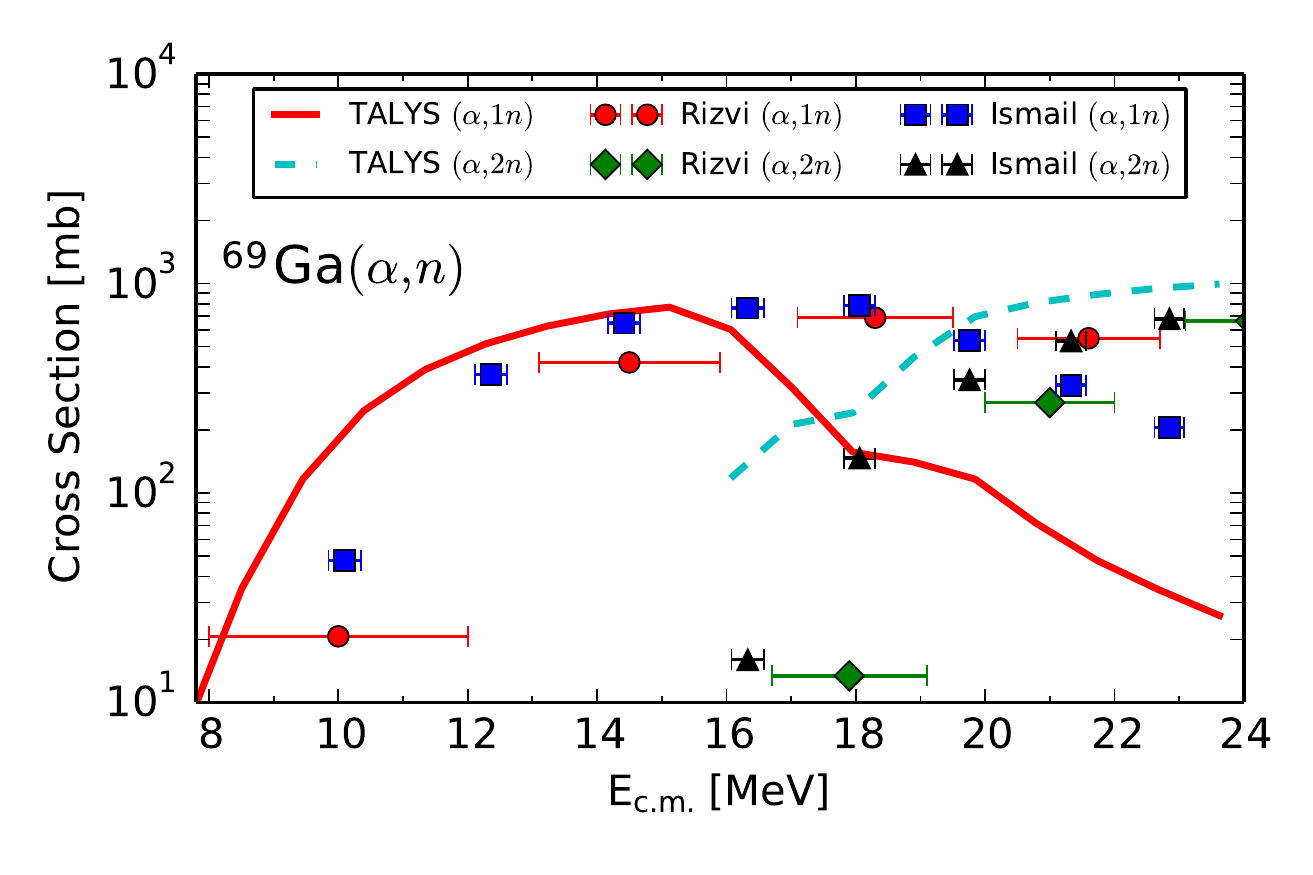}
\includegraphics[width=0.49\linewidth,angle=0]{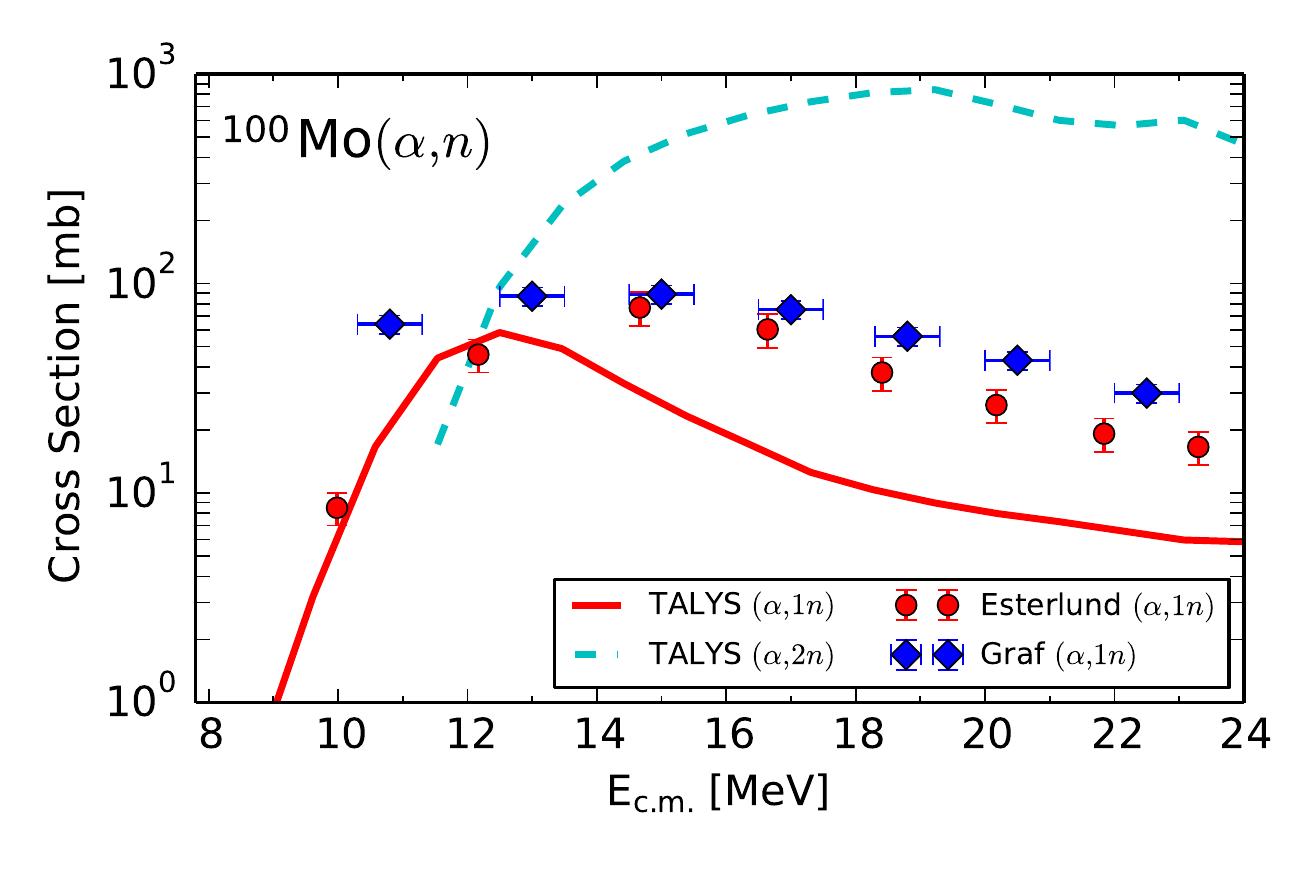}
\caption{Comparison of experimental and calculated cross sections for the reactions $^{69}$Ga$(\alpha,1n)$, $^{69}$Ga$(\alpha,2n)$ \cite{Rizvi.etal:1989} (left), and $^{100}$Mo$(\alpha,1n)$, $^{100}$Mo$(\alpha,2n)$ \cite{Ismail:1990} (right). The calculations were done with TALYS using the global alpha optical potential (GAOP).}
\label{fig:an_exp}
\end{figure}

The reaction flows (Fig.~\ref{fig:Flow}) and the temperature evolution of the averaged reaction time scales $\langle \tau_x \rangle$ (Fig.~\ref{fig:Timescale}) emphasize that $(\alpha,n)$ reactions are important for the weak r-process nucleosynthesis. Unfortunately, none of the relevant $(\alpha,n)$ reactions has been measured in the energy (temperature) range relevant for the astrophysical conditions discussed here. Therefore, one needs to use reaction codes, such as TALYS~\cite{Talys1.6} or NON-SMOKER~\cite{Rauscher.Thielemann:2000}, to calculate all the reaction rates entering into our nucleosynthesis network. Although these codes are based on the Hauser-Feshbach model \cite{Hauser.Feshbach:1952}, they can include important differences related to 1) intrinsic technical aspects and 2) nuclear physics inputs~\cite{Beard.etal:2014,Pereira.Montes:2016}. This arbitrariness in the treatment of the reaction leads to variations in the calculated rates. In a recent study~\cite{Pereira.Montes:2016}, the theoretical uncertainty of a selected group of $(\alpha,n)$ reactions relevant for the weak r-process was investigated. According to that work, $(\alpha,1n)$ is by far the most important $(\alpha,\times n)$ channel at temperatures relevant for the weak r-process between $T\simeq$ 2--5~GK (see also Ref.~\cite{Mohr:2016} for further information about $(\alpha,\times n)$ channels). Moreover, at these temperatures, the uncertainty in the calculated rates arises from the different models used to determine the alpha optical potential. In particular, the rates calculated using these different models can disagree by more than a factor 10 at  temperatures $T\simeq$ 2~GK~\cite{Pereira.Montes:2016,Mohr:2016}. This is illustrated in Fig.~\ref{fig:uncAOP} for the reactions
$^{69}$Ga$(\alpha,n)$$^{72}$As, $^{84}$Se$(\alpha,n)$$^{87}$Kr, $^{94}$Sr$(\alpha,n)$$^{97}$Zr, and $^{100}$Mo$(\alpha,n)$$^{103}$Ru. We include here also $^{69}$Ga and $^{100}$Mo because we use them later to compare experimental and theoretical cross sections (Fig.~\ref{fig:an_exp}).

While ambiguities in the alpha optical potentials govern the theoretical uncertainty at $T\simeq$ 2--5~GK, there are other aspects (e.g., level densities, binning of excitation energy) contributing to the theoretical uncertainty of the $(\alpha,n)$ reaction rates~\cite{Pereira.Montes:2016}. Although these aspects were found to have a rather limited impact in the calculated rates, larger discrepancies (of the order of $\sim$10) can be found between calculations and measurements at temperatures above the weak r-process regime, as shown in Fig.~\ref{fig:an_exp} for some $(\alpha,n)$ reactions. Notice that we compare with measurements of stable nuclei in absence of relevant experiments for the weak r-process where mainly unstable nuclei are involved.

In the light of the conclusions discussed in~\cite{Pereira.Montes:2016} and the results shown in Fig.~\ref{fig:an_exp}, it is reasonable to assume that the reliability of the calculated $(\alpha,n)$ rates is not better than a factor 10. In the present work, we investigate the sensitivity of abundances to \alphan uncertainties. We first calculate weak r-process abundances taking the \alphan TALYS~1.6 reaction rates calculated with the packet of models TALYS~1 (see Table II of Ref.~\cite{Pereira.Montes:2016}) except for masses, which were taken from the mass table of Ref.~\cite{Audi.etal:2003} if available, or from the FRDM mass model \cite{Moeller.etal:1995} otherwise. Then, we repeat the network calculations using the TALYS rates multiplied and divided by different factors, namely 5, 10, and 50.

% - - - - - - - - - - - - - - - - - - - - - - - - - - - - - - - - - - - - - - - -
\section{Impact of \alphan reaction rate uncertainties on the abundances} 
\label{sec:results}
\begin{figure}
\centering
\includegraphics[width=0.5\linewidth,angle=0]{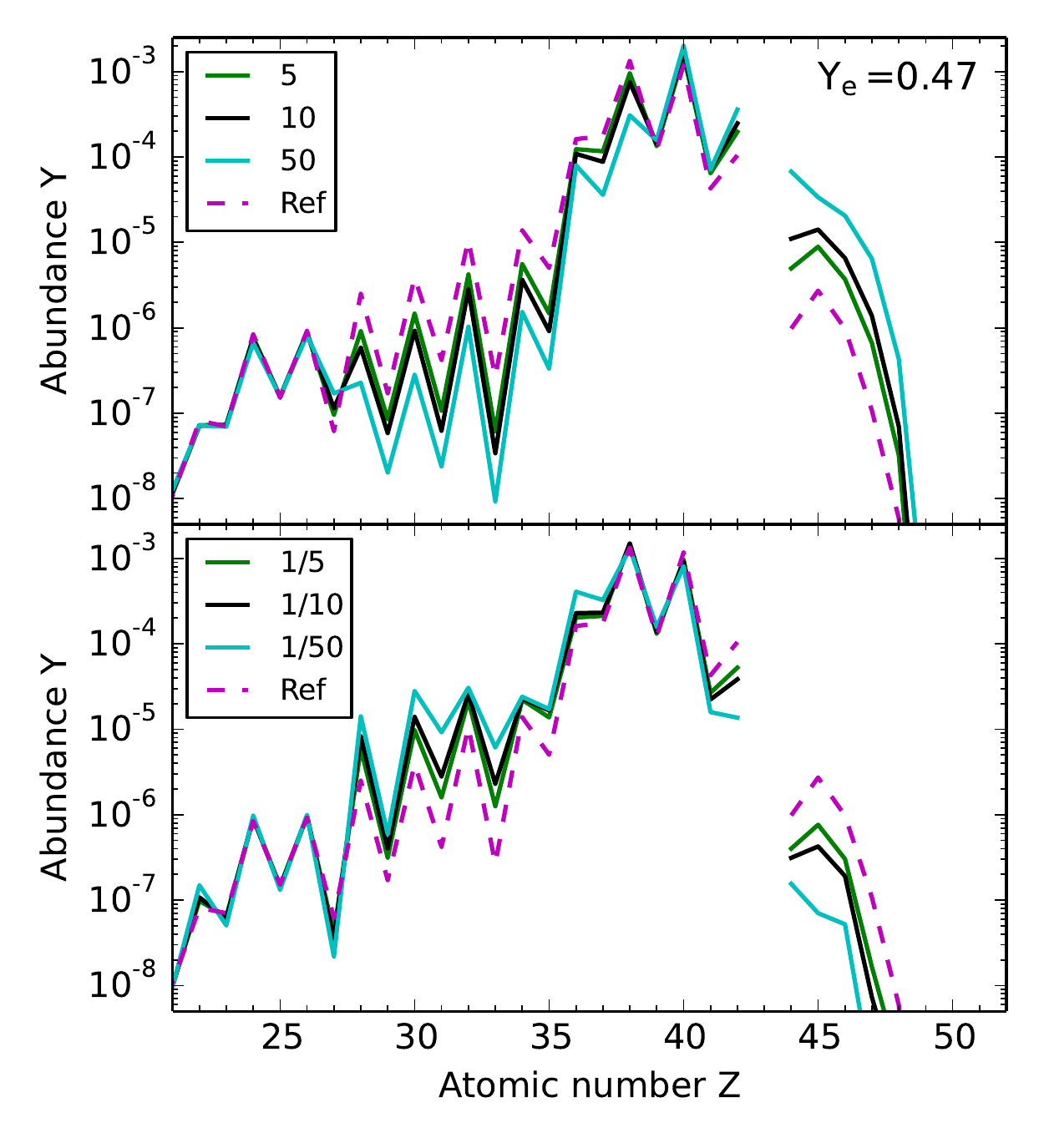}%
\includegraphics[width=0.5\linewidth,angle=0]{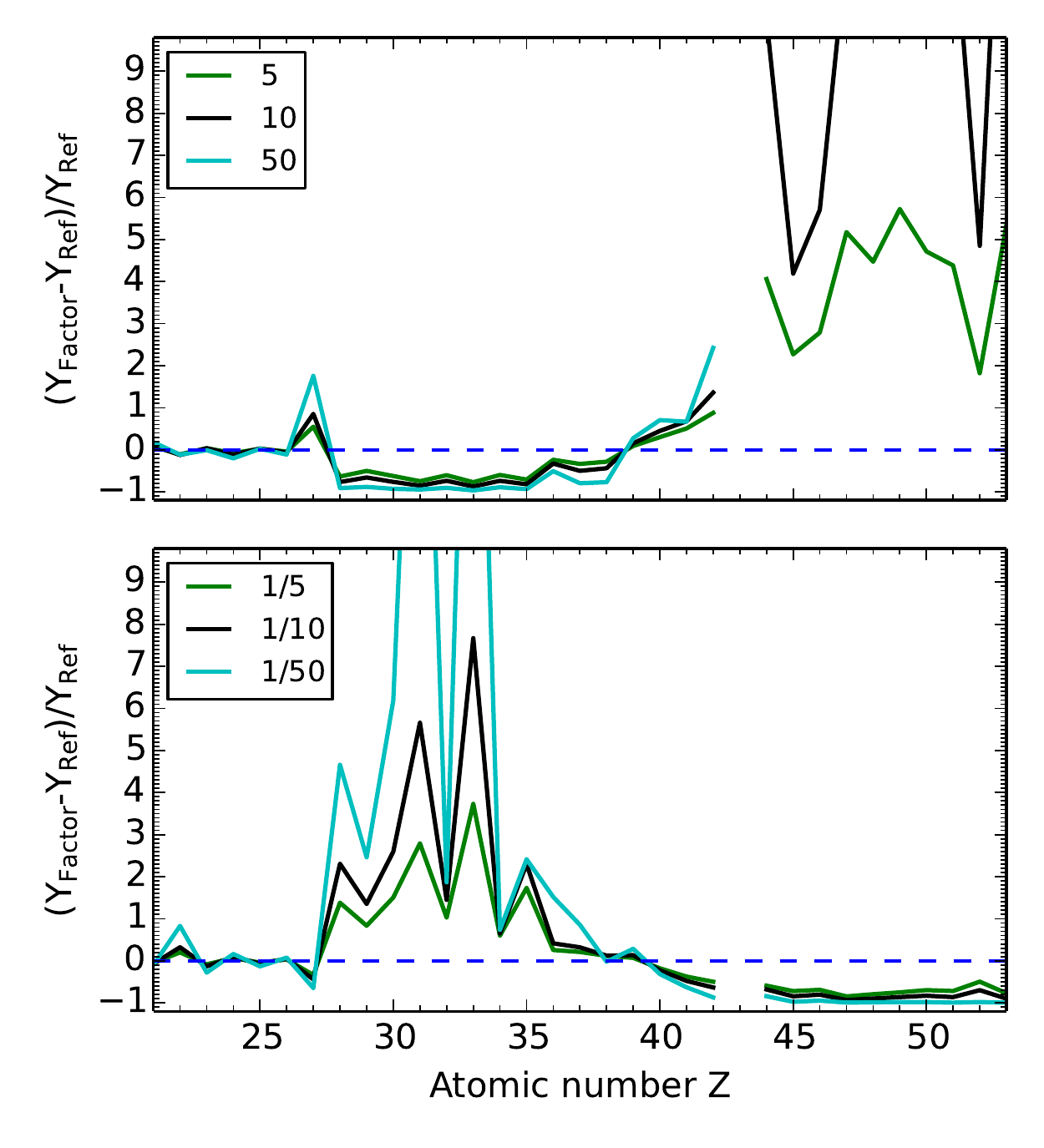}
\caption{Left panels: Elemental abundances when multiplying (upper panels) and dividing (lower panels) the $(\alpha,n)$ reaction 
  rates by factors 5, 10, and 50 for $Z=26-45$.
  The reference case corresponds to the 9~s trajectory with
  $Y_{e}=0.47$ with the original TALYS-calculated \alphan rates. Right panels: Relative changes of the abundances compared to the reference.}
\label{fig:uncabunYe47}
\end{figure}

% - - - - - - - - - - - - - - - - - - - - - - - - - - - - - - - - - - - - - - - -
We study the impact of \alphan reaction rate uncertainties on the abundances based on the trajectory introduced in Sect.~\ref{sec:nuc} for different \ye as discussed in Sect.~\ref{sec:astro_uncer}. We focus here on three initial electron fractions, i.e., \ye = 0.45, 0.47, and 0.49 and vary the \alphan reaction rates (and their inverse reactions) by constant factors (see Sect.~\ref{sec:an_uncer}). Figure~\ref{fig:uncabunYe47} (left panels) shows the final abundances for $Y_e=0.47$ when the $(\alpha,n)$ and $(n,\alpha)$ reaction rates are multiplied (upper panel) and divided (lower panel) by factors of 5, 10, and 50 for all isotopes between Fe and Rh. The relative abundance changes compared to the reference case are shown in the right panels of Fig.~\ref{fig:uncabunYe47}. All scaling factors have significant impact on the abundances. When the \alphan reactions become the fastest charged-particle reactions, the most abundant species are within the range $26 \lesssim Z \lesssim 40$ (Figs.~\ref{fig:Flow}-\ref{fig:Timescale}) and their abundances become thus sensitive to \alphan reaction rates within this range (see also \cite{Woosley.Hoffman:1992}).

When the \alphan rates are reduced (bottom panels, Fig.~\ref{fig:uncabunYe47}), less efficient alpha captures prevent nuclear matter from moving towards heavier nuclei, and thus the abundances stay higher between $27<Z<38$ compared to the reference case. This is clearly visible in the relative changes of abundances shown in the right bottom panel of Fig.~\ref{fig:uncabunYe47}. The abundances for $Z<38$ increase proportionally to the reduction factor used for the \alphan reactions. As less matter is moved beyond $Z<38$, the abundance for nuclei heavier than Zr decreases as indicated by the negative values of the relative changes. The opposite behavior is found for the increase of the rates (upper panels, Fig.~\ref{fig:uncabunYe47}), where the abundances of these nuclei become larger compared to the reference case. The relative changes (upper, right panel) are now negative for $Z<38$ indicating the reduction of abundances for such nuclei. The more efficient \alphan reactions move matter towards heavier nuclei as shown by the large and positive values (that go up to $\sim$ 100) for the relative abundance change.

Notice that the impact of $(\alpha,n)$ uncertainties on Y and Zr is relatively small (see right panels Fig.~\ref{fig:uncabunYe47}). For these conditions, there are almost no $(\alpha,n)$ reactions above Kr as shown by the reaction flows (Fig.~\ref{fig:Flow}). The $(\alpha,n)$ reactions on Kr isotopes are very important because they influence the abundances of the Sr isotopes directly. The flow from Sr isotopes towards heavier ones is mainly driven by \alphan reactions on Zr, $(p,n)$ reactions, and beta decays.

\begin{figure}
\centering
\includegraphics[width=0.5\linewidth,angle=0]{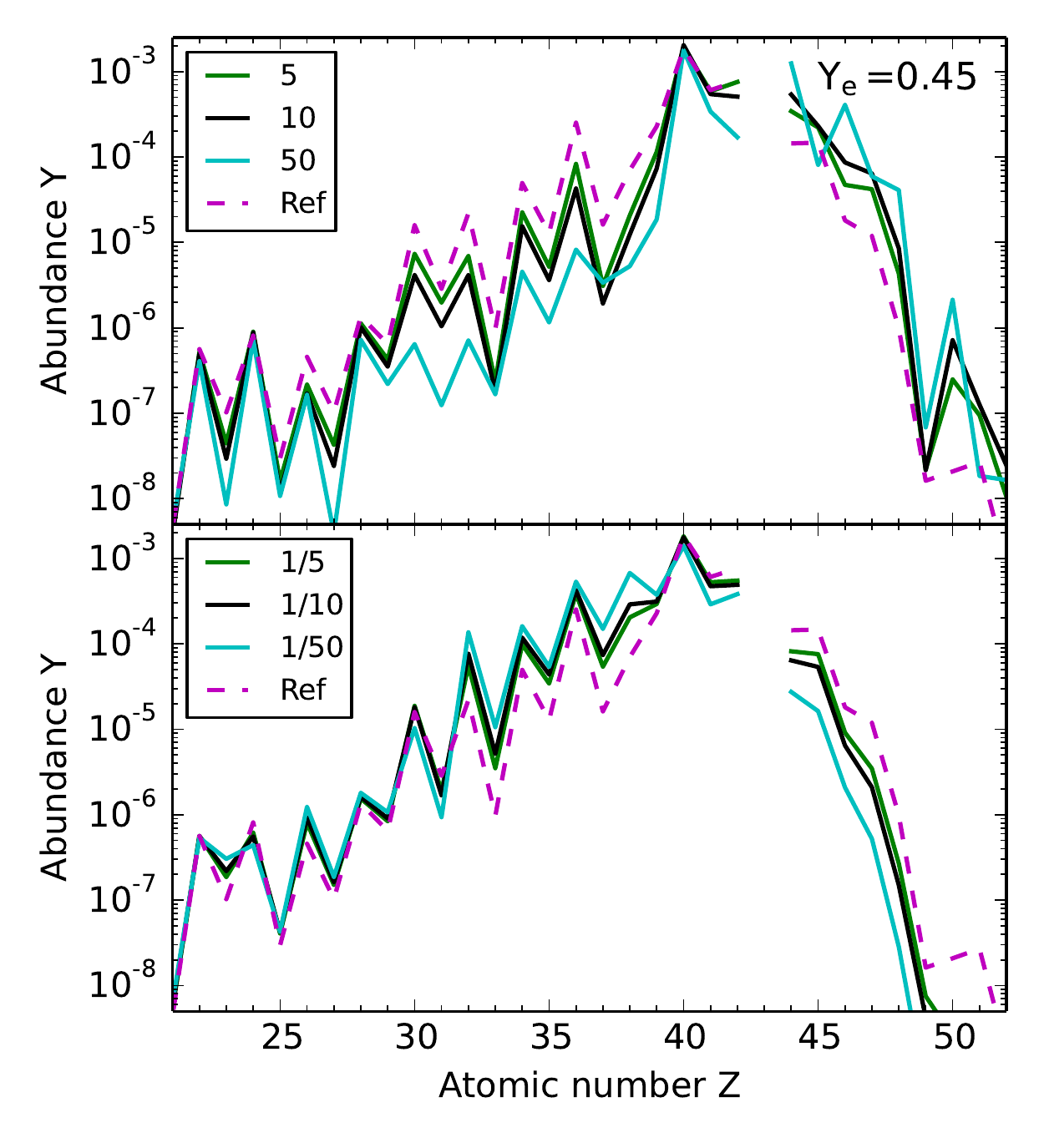}%
\includegraphics[width=0.5\linewidth,angle=0]{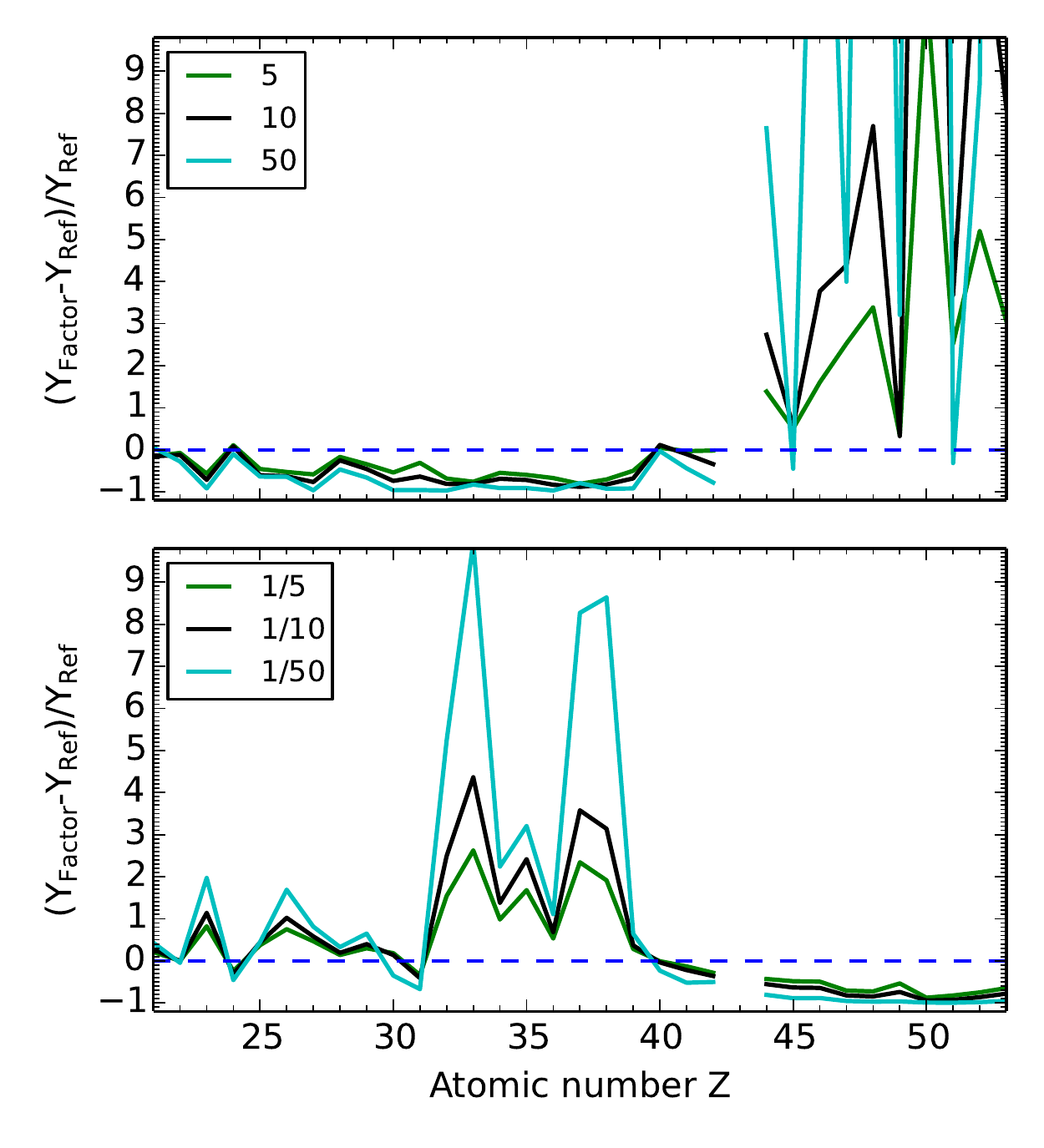}
\caption{Same as Fig.~\ref{fig:uncabunYe47} for $Y_{e}=0.45$. }
\label{fig:uncabunYe45}
\end{figure}

We have described the general trends of the abundances when varying the $(\alpha,n)$ reaction rates for the trajectory with $Y_{e} = 0.47$. The trend is the same for $Y_{e} = 0.45$ (Fig.~\ref{fig:uncabunYe45}). The main difference is the large variation in the abundance of Sr and the heaviest elements mainly due to $(\alpha,n)$ reactions on Rb and Sr isotopes. 
For both conditions $Y_{e} = 0.45$ and 0.47, there are also the changes in the abundances for $Z<26$ nuclei, whose \alphan rates are not scaled in our study. The change of these abundances is due to neutron captures that occur on all isotopes and are affected by the amount of neutrons available including the ones produce after \alphan reactions. For $Y_{e} = 0.49$, there is almost no effect when changing the $(\alpha,n)$ reactions. Here, the nucleosynthesis path moves along the valley of stability where $(p,\gamma)$, $(p,n)$ and few strong contributing $(\alpha,\gamma)$ reactions move matter towards heavy nuclei.

\begin{figure}
\centering
\includegraphics[width=0.75\linewidth,angle=0]{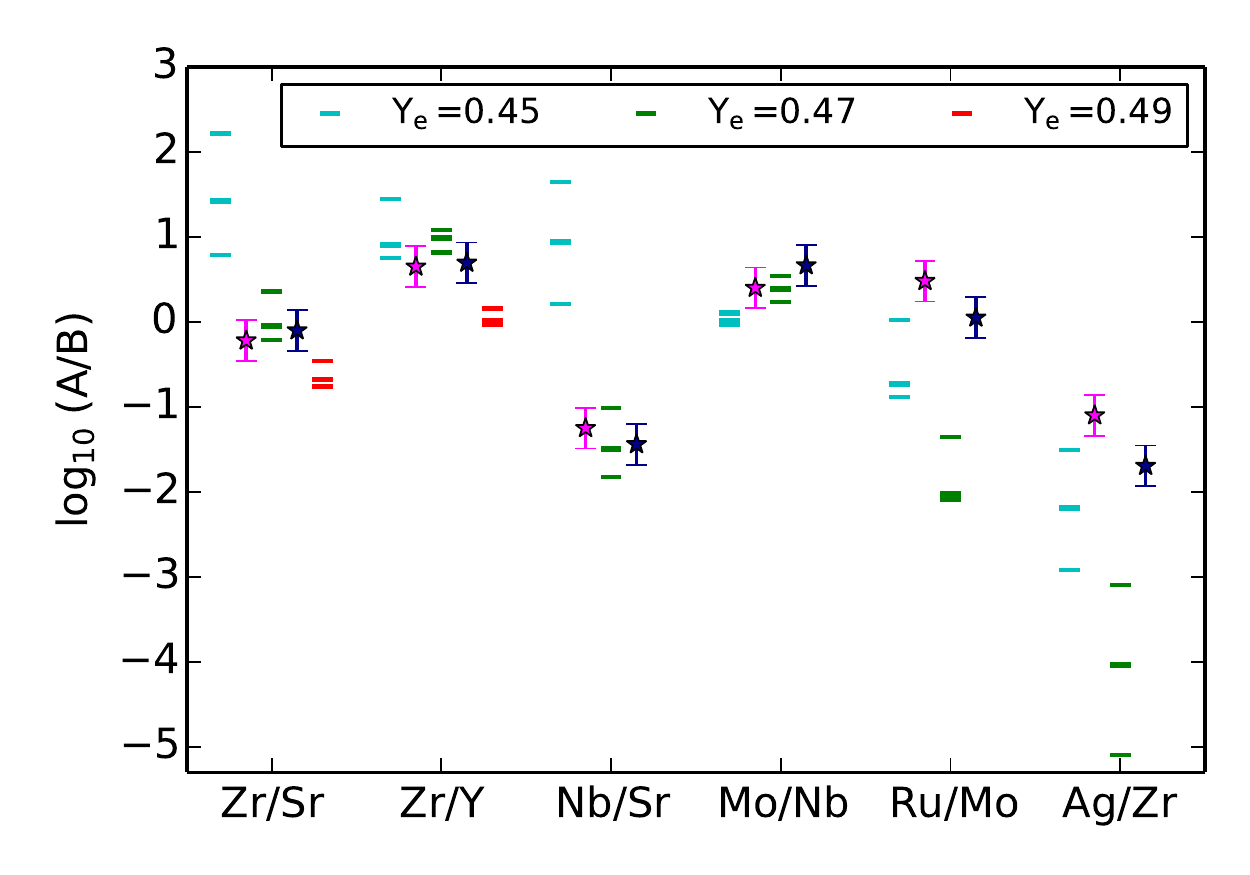}
\caption{Elemental abundance ratios compared to observations from averaged HD122563 and HD88609 (blue stars) and CS22892-052 (pink stars) for different electron fractions. The horizontal lines indicate the abundance ratio of the pair of elements: the thicker lines correspond to the reference TALYS rates, whereas the thin lines correspond to the reference rates scaled by factors of 10 and 0.1.}
\label{fig:observations}
\end{figure}

After studying the impact of different nucleosynthesis evolutions (various electron fractions) and of nuclear physics input due to \alphan reactions, we can combine these two uncertainties and compare to abundance observations of metal-poor stars.  In Fig.~\ref{fig:observations}, we show abundance ratios between pairs of lighter heavy element abundances in observations compared to our results. For the observations, we use metal-poor stars with typical r-process robust pattern (CS22892-052 \cite{Sneden.etal:2003}) and with low-enrichment of heavy elements (HD122563 and HD88609 \cite{Honda.etal:2007}) which are illustrated by pink and blue stars, respectively. For a detailed discussion about these representative patterns see~\cite{Hansen.etal:2014}. In Fig.~\ref{fig:observations}, the horizontal lines indicate the value of the abundance ratio between a given pair of elements A and B using the reference TALYS \alphan rates discussed in Sect.~\ref{sec:an_uncer} (thicker line) and varying this by factors 10 and 0.1. Different colors of the horizontal lines corresponds to different electron fractions (i.e., different nucleosynthesis paths). For \ye=0.45 there are large variations in the Zr/Sr, Nb/Sr, and Ag/Zr abundance ratios owing to the strong influence of \alphan rates on the abundances of Sr and Ag (Fig.~\ref{fig:uncabunYe45}). For ratios including heavier elements, i.e., Ru or Ag, there is more scatter due to the low abundances that rapidly drop for increasing proton number. In the case of $Y_{e}=0.47$, the variations in the Zr/Sr and Nb/Sr ratios can be explained by the impact of \alphan reactions on Sr and Nb abundances (Fig.~\ref{fig:uncabunYe47}). 
For \ye =0.49 we only show the Zr/Sr and Zr/Y abundance ratios because the final abundances do not reach nuclei heavier than $Z \sim 40$ (Fig.~\ref{fig:diffYe}). Due to the small influence of \alphan rates for \ye =0.49, the abundance ratios do not differ much when changing the rates. 
As Fig.~\ref{fig:observations} shows, it is clear that to make full use of the metal-poor observations when comparing to astrophysical models, nuclear physics uncertainties need to be reduced. In general, the various nucleosynthesis evolutions given by different values of $Y_{e}$ fail to reproduce the trend from Zr/Sr through Ag/Zr. Our aim here is not to find the exact astrophysical conditions that reproduce observations, but to combine astrophysics and nuclear physics uncertainties to show that both are critical to understand the production of lighter heavy elements. A complete study with broader variation of astrophysics and nuclear physics conditions will contribute to understand the post-explosion conditions using observations of ultra-metal poor stars, once experiments reduce the nuclear physics uncertainty.

\section{Summary}
\label{sec:summary}

In neutrino-driven winds formed after core-collapse supernovae, high initial temperatures keep the supersonically expanding matter in nuclear statistical equilibrium. After NSE, different reactions keep moving matter towards heavy nuclei and redistribute it. Therefore, it is critical to identify these reactions and reduce their uncertainty in order to be able to use observations of the oldest stars to constrain the astrophysical conditions found in the wind. In this paper, we have focused in slightly neutron-rich winds and found that \alphan reactions are key to understand the final abundances. Under these conditions a weak r-process takes place where the nucleosynthesis path moves close or along stability. There, \alphan reactions are faster than beta decays and thus they are responsible to keep matter moving towards heavy nuclei. We have described in detailed this nucleosynthesis process on the bases of the reaction flux and time scales. 

The astrophysical uncertainties have been analysed by varying the neutron-richness of the wind, i.e., \ye. This is not a complete study of all possible variations but allows us to investigate the impact of \alphan reactions under different conditions. In a forthcoming paper we will systematically vary the astrophysical conditions. In order to analyze the impact of \alphan reactions we have estimated first their uncertainties based on previous studies \cite{Pereira.Montes:2016,Mohr:2016}. These point to the alpha optical potential as the major source of uncertainty for the calculation of theoretical rates in the temperature range relevant for the weak r-process. Notice that there is no experimental information for \alphan reactions on unstable nuclei and the information for stable nuclei does not always agree with the theoretical predictions, as we have discussed here. 

After varying all \alphan reactions between Fe and Rh by multiplying and dividing them by factors 5, 10, and 50 we find that there is a significant impact on the final abundances. When the \alphan rates are reduced, less matter moves towards heavier nuclei, i.e., $Z>38$. This results in higher abundances below Sr compared to the case where no \alphan rates are varied. The opposite occurs for an increase of the rates, namely more matter moves up and the abundances are higher for $Z>38$. We find that this result is robust for different nucleosynthesis evolutions obtained by varying the electron fraction. Only when the electron fraction is very high, $Y_e=0.49$, the path moves along stability and $(\alpha, \gamma)$ reactions become very important while \alphan reactions have a negligible impact. Moreover, based on the fluxes, we were able to identify some key reactions, like \alphan on Kr isotopes or on $^{94}$Sr. However, a more detailed investigation would be necessary to identify all critical reactions. This is planned for future work including a Monte Carlo study for various astrophysical conditions.

With this first study, we have demonstrated that \alphan reactions are crucial to understand the production of lighter heavy elements up to Ag in neutrino-driven winds. Further effort is necessary to identify the most important reactions to be measured with the goal of reducing the nuclear physics uncertainties. This will open new and unique possibilities to use observations to constrain the extreme astrophysical condition where these elements are synthesized. Here we have summarized the importance of astrophysics and nuclear physics uncertainties in Fig.~\ref{fig:observations} where we have compared abundance ratios from observations to calculations including uncertainties.

%------------------------------------------------------------------------------------------------------
\ack
This work was funded by Helmholtz Young Investigator Group VH-NG-825,
Deutsche Forschungsgemeinschaft through SFB 1245 and the National Science
Foundation under Grant No. PHY-1430152 (JINA Center for the Evolution of the
Elements).

%%%%%%%%%%%%%%%%%%%%%%%%%%%%%%%%%%%%%%%%%%%%%%%%%%%%%%%%%%%%%%%
%\newpage
\section*{References}
%\bibliographystyle{jphysg}
%\bibliography{paper.bib}

\begin{thebibliography}{10}
\providecommand{\url}[1]{\texttt{#1}}
\providecommand{\urlprefix}{URL }
\providecommand{\eprint}[2][]{\url{#2}}

\bibitem{Meyer.etal:1992}
{Meyer} B~S, {Mathews} G~J et~al. 1992 \emph{\apj} \textbf{399} 656

\bibitem{Truran.Cowan:2000}
{Truran} J~W and {Cowan} J~J 2000 in {W~Hillebrandt \& E~M{\"u}ller}, editor,
  \emph{Nuclear Astrophysics, 2000}

\bibitem{Qian.Wasserburg:2007}
{Qian} Y and {Wasserburg} G~J 2007 \emph{\physrep} \textbf{442} 237

\bibitem{Pruet.etal:2006}
Pruet J, Hoffman R~D et~al. 2006 \emph{\apj} \textbf{644} 1028

\bibitem{Froehlich.etal:2006}
{Fr{\"o}hlich} C, {Mart{\'{\i}}nez-Pinedo} G et~al. 2006 \emph{\prl}
  \textbf{96} 142502

\bibitem{Wanajo:2006}
Wanajo S 2006 \emph{\apj} \textbf{647} 1323

\bibitem{Arnould.Goriely:2003}
{Arnould} M and {Goriely} S 2003 \emph{\prep} \textbf{384} 1

\bibitem{Cowan.Rose:1977}
{Cowan} J~J and {Rose} W~K 1977 \emph{\apj} \textbf{212} 149

\bibitem{Hampel.etal:2016}
{Hampel} M, {Stancliffe} R~J et~al. 2016 \emph{\apj} \textbf{831} 171

\bibitem{Jones.etal:2016}
{Jones} S, {Ritter} C et~al. 2016 \emph{\mnras} \textbf{455} 3848

\bibitem{Goriely.etal:2011}
{Goriely} S, {Bauswein} A et~al. 2011 \emph{\apjl} \textbf{738} L32

\bibitem{Korobkin.etal:2012}
{Korobkin} O, {Rosswog} S et~al. 2012 \emph{\mnras} \textbf{426} 1940

\bibitem{Wanajo.etal:2014}
{Wanajo} S, {Sekiguchi} Y et~al. 2014 \emph{\apjl} \textbf{789} L39

\bibitem{Ji.etal:2016}
{Ji} A~P, {Frebel} A et~al. 2016 \emph{\nat} \textbf{531} 610

\bibitem{Shibagaki.etal:2016}
{Shibagaki} S, {Kajino} T et~al. 2016 \emph{\apj} \textbf{816} 79

\bibitem{Nishimura.etal:2015}
{Nishimura} N, {Takiwaki} T et~al. 2015 \emph{\apj} \textbf{810} 109

\bibitem{Arcones.etal:2016}
{Arcones} A, {Bardayan} D~W et~al. 2016 \emph{White Paper on Nuclear
  Astrophysics} \eprint{[arXiv:1603.02213]}

\bibitem{Mumpower.etal:2016}
Mumpower M, Surman R et~al. 2016 \emph{Progress in Particle and Nuclear
  Physics} \textbf{86} 86–126 ISSN 0146-6410

\bibitem{Martin.etal:2016}
{Martin} D, {Arcones} A et~al. 2016 \emph{Physical Review Letters} \textbf{116}
  121101

\bibitem{Hansen.etal:2014}
{Hansen} C~J, {Montes} F et~al. 2014 \emph{\apj} \textbf{797} 123

\bibitem{Qian.Wasserburg:2008}
{Qian} Y~Z and {Wasserburg} G~J 2008 \emph{\apj} \textbf{687} 272

\bibitem{Travaglio.etal:2004}
{Travaglio} C, {Gallino} R et~al. 2004 \emph{\apj} \textbf{601} 864

\bibitem{Montes.etal:2007}
{Montes} F, {Beers} T~C et~al. 2007 \emph{\apj} \textbf{671} 1685

\bibitem{Arcones.Montes:2011}
{Arcones} A and {Montes} F 2011 \emph{\apj} \textbf{731} 5

\bibitem{Hoffman.etal:1997}
{Hoffman} R~D, {Woosley} S~E et~al. 1997 \emph{\apj} \textbf{482} 951

\bibitem{Huedepohl.etal:2010}
{H{\"u}depohl} L, {M{\"u}ller} B et~al. 2010 \emph{\prl} \textbf{104} 251101

\bibitem{Arcones.Thielemann:2013}
{Arcones} A and {Thielemann} F~K 2013 \emph{Journal of Physics G Nuclear
  Physics} \textbf{40} 013201

\bibitem{Roberts.etal:2012}
{Roberts} L~F, {Reddy} S et~al. 2012 \emph{\prc} \textbf{86} 065803
  \eprint{1205.4066}

\bibitem{MartinezPinedo.etal:2012}
{Mart{\'{\i}}nez-Pinedo} G, {Fischer} T et~al. 2012 \emph{Physical Review
  Letters} \textbf{109} 251104

\bibitem{Arcones.Bliss:2014}
{Arcones} A and {Bliss} J 2014 \emph{Journal of Physics G Nuclear Physics}
  \textbf{41} 044005

\bibitem{Woosley.Hoffman:1992}
{Woosley} S~E and {Hoffman} R~D 1992 \emph{\apj} \textbf{395} 202

\bibitem{Sasaqui.etal:2005}
{Sasaqui} T, {Kajino} T et~al. 2005 \emph{\apj} \textbf{634} 1173

\bibitem{Pereira.Montes:2016}
{Pereira} J and {Montes} F 2016 \emph{Phys. Rev. C} \textbf{93} 034611

\bibitem{Mohr:2016}
Mohr P 2016 \emph{Phys. Rev. C} \textbf{94} 035801

\bibitem{Hauser.Feshbach:1952}
{Hauser} W and {Feshbach} H 1952 \emph{Physical Review} \textbf{87} 366

\bibitem{Talys1.6}
{Koning} A~J, {Hilaire} S et~al. 2013 {Talys 1.6 user manual,
  \url{http://www.talys.eu/fileadmin/talys/user/docs/talys1.6.pdf}}

\bibitem{Winteler:2012}
{Winteler} C 2012 \emph{{}} Ph.D. thesis Univ. Basel, CH

\bibitem{Winteler.etal:2012}
{Winteler} C, {K{\"a}ppeli} R et~al. 2012 \emph{\apjl} \textbf{750} L22

\bibitem{Cyburt.etal:2010}
{Cyburt} R~H, {Amthor} A~M et~al. 2010 \emph{\apjs} \textbf{189} 240

\bibitem{Qian.Woosley:1996}
{Qian} Y~Z and {Woosley} S~E 1996 \emph{\apj} \textbf{471} 331

\bibitem{Takahashi.etal:1994}
{Takahashi} K, {Witti} J et~al. 1994 \emph{\aap} \textbf{286}

\bibitem{Arcones.etal:2007}
{Arcones} A, {Janka} H~T et~al. 2007 \emph{\aap} \textbf{467} 1227

\bibitem{Fowler.etal:1967}
Fowler W~A, Caughlan G~R et~al. 1967 \emph{\araa} \textbf{5} 525–570

\bibitem{Arcones.Janka:2011}
{Arcones} A and {Janka} H~T 2011 \emph{\aap} \textbf{526} A160

\bibitem{Otsuki.etal:2000}
{Otsuki} K, {Tagoshi} H et~al. 2000 \emph{\apj} \textbf{533} 424

\bibitem{Wanajo:2013}
{Wanajo} S 2013 \emph{\apjl} \textbf{770} L22

\bibitem{Wanajo.etal:2011}
{Wanajo} S, {Janka} H~T et~al. 2011 \emph{\apjl} \textbf{726} L15

\bibitem{Sneden.etal:2008}
{Sneden} C, {Cowan} J~J et~al. 2008 \emph{\araa} \textbf{46} 241

\bibitem{Hansen.etal:2013}
{Hansen} C~J, {Bergemann} M et~al. 2013 \emph{\aap} \textbf{551} A57

\bibitem{Roederer.etal:2014b}
{Roederer} I~U, {Preston} G~W et~al. 2014 \emph{\apj} \textbf{784} 158

\bibitem{Rauscher.Thielemann:2000}
{Rauscher} T and {Thielemann} F~K 2000 \emph{\adndt} \textbf{75} 1

\bibitem{Beard.etal:2014}
Beard M, Uberseder E et~al. 2014 \emph{Phys. Rev. C} \textbf{90} 034619

\bibitem{Watanabe:1958}
{Watanabe} S 1958 \emph{\npa} \textbf{8} 484

\bibitem{McFadden.Sachtler:1966}
McFadden L and Satchler G~R 1966 \emph{Nuclear Physics} \textbf{84} 177

\bibitem{Demetriou.etal:2002}
{Demetriou} P, {Grama} C et~al. 2002 \emph{Nuclear Physics A} \textbf{707} 253

\bibitem{Audi.etal:2003}
Audi G, Bersillon O et~al. 2003 \emph{Nuclear Physics A} \textbf{729} 3–128
  ISSN 0375-9474

\bibitem{Moeller.etal:1995}
{M{\"o}ller} P, {Nix} J~R et~al. 1995 \emph{Atomic Data and Nuclear Data
  Tables} \textbf{59} 185

\bibitem{Rizvi.etal:1989}
{Rizvi} I~A, {Bhardwaj} M~K et~al. 1989 \emph{Canadian Journal of Physics}
  \textbf{67} 870

\bibitem{Ismail:1990}
{Ismail} M 1990 \emph{\prc} \textbf{41} 87

\bibitem{Sneden.etal:2003}
{Sneden} C, {Cowan} J~J et~al. 2003 \emph{Nuclear Physics A} \textbf{718} 29

\bibitem{Honda.etal:2007}
{Honda} S, {Aoki} W et~al. 2007 \emph{\apj} \textbf{666} 1189
  \eprint{0705.3975}

\end{thebibliography}

\end{document}